\documentclass[a4paper,11pt,fleqn]{article}
\usepackage{amsfonts,latexsym,amstext}

\def\nn{\nonumber}
\def\non{\nonumber\\}
\def\be{\begin{equation}}
\def\ee{\end{equation}}
\def\ben{\begin{displaymath}}
\def\een{\end{displaymath}}
\def\ba{\begin{eqnarray}}
\def\ea{\end{eqnarray}}

\def\a{\alpha}
\def\ua{{\underline\alpha}}
\def\b{\beta}
\def\ub{{\underline\beta}}
\def\D{\Delta}
\def\d{\delta}
\def\e{\varepsilon}

\def\g{\gamma}
\def\G{\Gamma}

\def\L{{\tilde\xi}}
\def\m\mu 
\def\n{\nu}

\def\e{\epsilon}

\def\moth{\mathsurround=0pt}
\newdimen\zo \zo=0pt

\def\tick{\leaders\hrule height 0.5ex depth 0pt \hskip 0.5pt}
\def\upboxfill{$\moth \setbox\zo\hbox{\tick}%
  \hskip 3pt\hbox to 0pt{$\tick$\hss}\hrulefill \hbox to 7.5pt{$\tick$\hss}$}
\def\underbox#1{\offinterlineskip{\mathord{\mathop{\vtop{\moth\ialign{##\crcr
      $\hfil\displaystyle{#1}\hfil$\crcr\noalign{}
      {\upboxfill}\crcr\noalign{}}}}\limits}}}
\def\dtick{\leaders\hrule height .34pt depth 0.5ex \hskip 0.5pt}
\def\downboxfill{$\moth \setbox\zo\hbox{\dtick}%
  \hskip 2pt\hbox to 0pt{$\dtick$\hss}\hrulefill \hbox to 2pt{$\dtick$\hss}$}
\def\overbox#1{\mathop{\vbox{\moth\ialign{##\crcr\noalign{}
\downboxfill\crcr\noalign{\vskip 1pt\nointerlineskip}
      $\hfil\displaystyle{#1}\hfil$\crcr}}}\limits}

\def\undersym#1{\underbox{{}#1}}
\def\oversym#1{\overbox{{}#1}}

\def\cA{{\cal A}}
\def\cB{{\cal B}}
\def\cC{{\cal C}}
\def\cD{{\cal D}}
\def\cE{{\cal E}}

\def\cG{{\cal G}}

\def\cM{{\cal M}}
\def\cN{{\cal N}}

\def\cP{{\cal P}}

\def\cV{{\cal V}}
\def\cX{{\cal X}}

\def\Zt{{\tilde{E}}}

\def\R{\mathbb{R}}

\def\E{E_8}

\def\EE{E_{8(8)}}
\def\9{E_9}
\def\0{E_{10}}
\def\SO{{SO}(16)}

\def\la{\label}
\def\ci{\cite}

\def\Ref#1{(\ref{#1})}

\def\ft#1#2{{\textstyle {\frac{#1}{#2}} }}
\def\8{\infty}
\def\p{\partial}

\def\tr{{\rm Tr \,}}
\def\Tr{{\rm Tr \,}}
\def\ra{\rightarrow}
\def\lra{\longrightarrow}

\def\16{N\!=\! 16}
\def\216{d\!=\!2 , N\!=\!16}
\def\316{d\!=\!3 , N\!=\!16}
\def\48{d\!=\!4 , N\!=\!8}

\makeatletter
\@addtoreset{equation}{section}
\makeatother
\renewcommand{\theequation}{\thesection.\arabic{equation}}

\begin{document}
\thispagestyle{empty}

\begin{flushright}
AEI-2000-009\\
LPTENS-00/24\\
hep-th/0006034
\end{flushright}
\renewcommand{\thefootnote}{\fnsymbol{footnote}}

\vspace*{0.1cm}
\begin{center}
\mathversion{bold}
{\bf\Large An exceptional geometry for $d=11$ supergravity?}
\bigskip\bigskip
\mathversion{normal}

{\bf\large K.~Koepsell, H.~Nicolai\medskip\\ }
{\large Max-Planck-Institut f{\"u}r Gravitationsphysik,\\
  Albert-Einstein-Institut,\\
  M\"uhlenberg 1, D-14476 Potsdam, Germany}~\footnote{ Supported by EU
  contract ERBFMRX-CT96-0012.}\\

\smallskip {\small E-mail: koepsell@aei-potsdam.mpg.de,
  nicolai@aei-potsdam.mpg.de} \bigskip

\setcounter{footnote}{0} {\bf\large H.~Samtleben\medskip\\ }
{\large Laboratoire de Physique Th{\'e}orique\\
  de l'Ecole Normale Sup{\'e}rieure,\\
  24 Rue Lhomond, 75231 Paris C{\'e}dex 05, France}~\footnotemark$^,\!$
\setcounter{footnote}{2}\footnote{UMR 8548: Unit{\'e} 
  Mixte du Centre National de la Recherche Scientifique, 
  et de l'{\'E}cole Normale Sup{\'e}rieure. }\\

\smallskip {\small E-mail:
henning@lpt.ens.fr\medskip} 
\end{center}
\renewcommand{\thefootnote}{\arabic{footnote}}
\setcounter{footnote}{0}
\bigskip
\medskip
\begin{abstract}
We analyze the algebraic constraints of the generalized vielbein 
in $SO(1,2)\times \SO$ invariant $d=11$ supergravity, and show that 
the bosonic degrees of freedom of $d=11$ supergravity, which become
the physical ones upon reduction to $d=3$, can be assembled
into an $\EE$-valued vielbein already in eleven dimensions.
A crucial role in the construction is played by the 
maximal nilpotent commuting subalgebra of $\EE$, of dimension 36, 
suggesting a partial unification of general coordinate and tensor 
gauge transformations.

\end{abstract}

\renewcommand{\thefootnote}{\arabic{footnote}}
\vfill
\leftline{{\sc June 2000}}

\setcounter{footnote}{0}
\newpage

\setcounter{page}{1}

\section{Introduction.} 

One of the most remarkable properties of maximal ($d=11$) supergravity
\ci{CrJuSc78} is the emergence of hidden symmetries of exceptional type
in the reduction to lower dimensions \ci{CreJul79}. Some time ago, it was
shown \ci{WitNic86,Nico87} that, already in eleven dimensions, this
theory permits reformulations where the tangent space symmetry 
$SO(1,10)$ is replaced by the local symmetries that would arise in the 
reduction to four and three dimensions, i.e. $SO(1,3)\times SU(8)$ 
and $SO(1,2)\times\SO$, respectively. The key point here is, of course, 
that this construction works without any assumptions restricting the 
dependence of theory on the coordinates, so these symmetries already
exist in eleven dimensions. Besides its general interest, this result
plays a pivotal role in establishing the consistency of the Kaluza Klein 
reduction of $d=11$ supergravity in various non-trivial backgrounds 
\ci{WitNic84,WitNic87}. In this article, we will concentrate on the 
$SO(1,2)\times\SO$ version of \ci{Nico87}, which exhibits special features 
(especially ``maximal unification'' of symmetries), and which in the reduction
to three dimensions yields directly maximal $N=16$ supergravity\footnote{The 
$\EE$ invariance of maximal $N=16$ supergravity in three dimensions 
was originally shown in \ci{Juli83}, while its complete Lagrangian 
and supersymmetry transformations were derived in \ci{MarSch83}. The latter 
follow directly by dimensional reduction of the variations presented 
in \cite{Nico87}. See also \ci{Mizo97,CJLP98} for a more recent treatment.}.

The equivalence of the different versions of $d=11$ supergravity is 
established at the level of the equations of motion by making special 
gauge choices, and does not extend off shell because the new versions 
mix equations of motion and Bianchi identities of the original theory. 
As shown in \ci{WitNic86,Nico87}, the bosonic fields, that would become 
scalar matter fields in the dimensional reduction, can be assigned to 
representations of the hidden global symmetry groups $E_{7(7)}$ and $\EE$, 
respectively. For this purpose it was necessary to fuse the bosonic
fields into new objects, christened ``generalized vielbeine'':
these are soldering forms with upper world indices running over the 
internal dimensions, and lower indices belonging to the $\bf 56$ and 
$\bf 248$ representations of $E_{7(7)}$ and $\EE$, respectively.
The generalized vielbeine are subject to algebraic constraints which 
follow from their explicit expressions in terms of the vielbein 
of $d=11$ supergravity. For the $SO(1,2)\times\SO$ version of \cite{Nico87},
there exists a differential constraint in addition, which is a $d=11$ 
variant of the duality constraint by which vector fields are converted 
into scalars in three dimensions. However, it was not clear from 
previous work how to solve these constraints (apart from the 
torus reduction, explicit solutions are only known for the $S^7$ 
compactification of $d=11$ supergravity \cite{WitNic87}), and how 
to recover the correct counting of bosonic physical degrees of freedom. 

An essential new element of the present work in comparison with 
previous results is our treatment of the 3-form potential $A_{MNP}$. 
Whereas in \ci{WitNic86,Nico87} this field appeared only via its field 
strength $F_{MNPQ} = 24 \p_{[M} A_{NPQ]}$ inside the $E_{7(7)}$ 
and $\EE$ connections given there, part of it here is merged into an
enlarged generalized vielbein. As a consequence, the latter now also 
transforms under tensor gauge transformations, suggesting a 
partial unification of coordinate and tensor gauge transformations. 
We show that the generalized vielbein is actually part of a 
full $\EE$ matrix $\cV$ (a ``248-bein''), which now lives in 
eleven dimensions, and which incorporates the bosonic degrees of 
freedom of $d=11$ supergravity, with the exception of the dreibein 
remaining from the 3+8 split (which does not propagate in the 
dimensionally reduced theory). Moreover, we present evidence for the 
existence of an ``exceptional geometry'' for $d=11$ supergravity 
by displaying the action of the combined internal coordinate 
and tensor gauge transformations on this 248-bein.
In deriving these results, we will make crucial use of certain 
special properties of the exceptional Lie algebra $\EE$, in particular 
the existence of a maximal nilpotent abelian subalgebra of  
dimension 36, which is unique up to conjugation \ci{Malc45},
and whose importance was recently emphasized in \ci{CJLP98}.

The present paper deals mainly with the algebraic relations
obeyed by the generalized vielbein, and their solution. The 
corresponding differential relations will be discussed elsewhere.
We believe that our results constitute further evidence for
a hidden $\EE$ structure of $d=11$ supergravity, but there remain
a number of open problems that must be dealt with; these include
in particular the proper treatment of the tensor components 
$B_{\mu\nu m}$ and $B_{\mu\nu\rho}$, and the construction of 
an invariant action in terms of the 248-bein $\cV$ in eleven
dimensions.

\mathversion{bold}
\section{$SO(1,2)\times \SO$ invariant $d=11$ supergravity} 
\mathversion{normal}
We will first review $SO(1,2)\times SO(16)$ formulation of $d=11$ 
supergravity, referring readers to \ci{Nico87} for further details. 
Our conventions concerning $\EE$ as well as its $\SO$ and $SL(8,\R)$ 
decompositions, which played an important role also in \ci{CJLP98},
are summarized in two appendices. 

$SO(1,2)\times \SO$ invariant $d=11$ supergravity \ci{Nico87} is 
derived from the original version of \ci{CrJuSc78} by first splitting 
up the fields in a way that would be appropriate for the reduction
to three dimensions, but without dropping the dependence on any
coordinates, and then reassembling the pieces into new objects 
transforming under local $SO(1,2)\times \SO$. Hence we are still 
dealing with $d=11$ supergravity, albeit in a very different guise. 
This is achieved by first breaking the original tangent space 
symmetry $SO(1,10)$ down to $SO(1,2)\times SO(8)$ by a partial
gauge choice for the elfbein, and then reenlarging it to 
$SO(1,2)\times SO(16)$ by the introduction of new gauge degrees 
of freedom. The construction thus requires a 3+8 split of the 
$d=11$ coordinates and tensor indices. The main task then is to 
identify the proper $SO(1,2)\times\SO$ covariant fields and to verify 
that all supersymmetry variations as well as the equations 
of motion can be entirely expressed in terms of the new fields.

In a first step one thus brings the elfbein into triangular form by 
(partial) use of local $SO(1,10)$ Lorentz invariance
\be
{E_M}^A =
\left(\begin{array}{cc}
\D^{-1}e_\mu{}^\ua& B_\mu{}^m {e_m}^a\\
0&{e_m}^a
\end{array}\right) \: , \qquad  \D := \det e_m{}^a , \la{vielbein}
\ee
Here curved $d=11$ indices decompose as $M=(\mu,m), N=(\nu,n),\ldots$ with
$\mu,\nu,\ldots=0,1,2$ and $m,n,\ldots=3,\ldots,10$, and the associated 
flat indices are denoted by $\ua,\ub,\ldots$ and $a,b,\ldots$, 
respectively\footnote{We hope that no confusion results from 
our double usage of $A,B,\dots$ as both $SO(1,10)$ vector and 
$\SO$ spinor indices. It should always be clear from the context
which is meant.} (as in \ci{Nico87} the $SO(1,2)$ 
indices $\ua,\ub,\dots $ are underlined to distinguish 
them from the $SO(8)$ spinor indices to be used below). 
The partially gauge fixed elfbein, whose form is preserved by
the $SO(1,2)\times SO(8)$ subgroup of $SO(1,10)$, thus contains 
the Weyl rescaled dreibein $e_\mu{}^\a$, the Kaluza-Klein vector 
$B_\mu{}^m$ and the achtbein ${e_m}^a$ yielding the scalar degrees 
of freedom living in the coset $GL(8,\R)/SO(8)$. 

The remaining bosonic degrees of freedom reside in the 3-index 
field $A_{MNP}$, which gives rise to various scalar and tensor 
fields upon performing a 3+8 split of the indices. First of all, 
there are 56 scalars $A_{mnp}$ and 28 vector fields
\be
B_{\mu mn} := A_{\mu mn} - {B_\mu}^p A_{mnp}
\ee
If one were to reduce to three dimensions, the 8+28 vector fields 
$B_\mu{}^m$ and $B_{\mu mn}$ would be converted to 36 scalar
degrees of freedom by means of a duality transformation. Here, they
will be kept together with their dual scalars contained in the 
generalized vielbein, to which they are related by a nonlinear analog of 
the (linear) duality constraint of the reduced $d=3$ supergravity. 

In addition, $A_{MNP}$ gives rise to (always in the 3+8 split)
\ba
B_{\mu\nu p} &:=& A_{\mu\nu p} - 2 {B_{[\nu}}^n A_{\mu]np} +
                   {B_{\mu}}^m {B_{\nu}}^n A_{mnp} \nn \\
B_{\mu\nu\rho} &:=& A_{\mu\nu\rho} - 3 {B_{[\mu}}^m A_{\nu\rho]m} 
             + 3 {B_{[\mu}}^m {B_{\nu}}^n A_{\rho]mn} \nn \\
      && - {B_\mu}^m {B_\nu}^n {B_\rho}^p A_{mnp}
\ea
These fields are subject to the tensor gauge transformations
\be
\d A_{MNP} = 3 \p_{[M} \xi_{NP]}\
\ee
Under these, we have
\ba
\d B_{\mu mn} &=& \cD_\mu \xi_{mn} + 2 \p_{[m} {B_\mu}^p \xi_{n]p}
                  + 2 \p_{[m} \L_{n]\mu} \nn \\
\d B_{\mu\nu m} &=& \p_m \L_{\mu\nu} + 2 \p_m {B_{[\mu}}^n \L_{n\nu]}
     +  2 \cD_{[\mu} \L_{\nu ]m} + {{\cB}_{\mu\nu}}^n \xi_{nm} \nn \\
\d B_{\mu\nu\rho} &=& 3 \cD_{[\mu} \L_{\nu\rho]} 
        - 3 {{\cB}_{[\mu \nu}}^m \L_{\rho]m}
\ea
where
\be
\cD_\mu := \p_\mu - {B_\mu}^m \p_m
\ee
and
\be\la{Bmunu}
{{\cB}_{\mu\nu}}^m := \cD_\mu {B_\nu}^m - \cD_\nu {B_\mu}^m
\ee
The parameters $\L_{\mu\nu}$ and $\L_{\mu m}$ are defined by
\ba
\L_{\mu\nu} &:=& \xi_{\mu\nu} + 2 {B_{[\mu}}^m \xi_{\nu] m}
                  +  {B_{\mu}}^m {B_{\nu}}^n \xi_{mn} \nn \\
\L_{\mu m} &:=& \xi_{\mu m} - {B_{\mu}}^n \xi_{nm}
\ea
It is easy to see that in the dimensionally reduced theory,
one can make use of the parameter components $\L_{\mu m}$
and $\L_{\mu\nu}$ to set $B_{\mu\nu m} = B_{\mu\nu\rho} =0$.
Since we have not been able so far to cast the supersymmetry
variations of these components into a completely $\SO$ covariant form,
this gauge would also be very convenient in the present setting. 
However, there appears to be an obstacle to this gauge choice
if the full $d=11$ coordinate dependence is retained.

To identify the proper $\SO$ covariant bosonic fields, we must first
explain how to rewrite the fermion fields. The $d=11$ gravitino 
$\Psi_A\equiv (\Psi_\ua ,\Psi_a)$ has 32 spinor components, which 
split as ${\bf 2}\otimes ({\bf 8}_s \oplus {\bf 8}_c)$ under the 
$SO(1,2)\times SO(8)$ subgroup of $SO(1,10)$. Suppressing $SO(1,2)$ spinor 
indices, we then assign the resulting fields to the $SO(1,2)\times\SO$ 
fields $\psi^I_\mu$ and $\chi^{\dot A}$ via the following 
prescription \ci{Nico87}
\be
\psi_\mu^I:=\left\{
               \begin{array}{ll}
\D^{-1/2}{e_\mu}^\ua \left( \Psi_{\ua \a} + \g_\ua \G^a_{\a\dot\b}
            \Psi_{a\dot\b}\right) &\text{ if } I =  \a \\
\D^{-1/2}{e_\mu}^\ua \left( \Psi_{\ua \dot\a}- \g_\ua \G^a_{\dot\a\b} 
            \Psi_{a\b}\right) &\text{ if } I = \dot \a 
\end{array}\right. \la{gravitino}
\ee
and
\be
\chi^{\dot A}:=\left\{
               \begin{array}{ll}
\D^{-1/2} \left( \G^b\G^a \right)_{\a\b} \Psi_{b\b}
            &\text{ if } \dot A =  (a \a ) \\
- \D^{-1/2} \left( \G^b\G^a \right)_{\dot\a \dot \b} \Psi_{b\dot\b}
            &\text{ if } \dot A =  (\dot\a a) 
\end{array}\right. \la{chi}
\ee
where $I$ and $\dot A$ are $\SO$ vector and (conjugate) spinor indices,
respectively (see appendix B, and in particular \Ref{SO8dec} for the 
relevant $SO(8)$ decompositions).

The physical bosonic degrees of freedom, which correspond to the 
propagating 128 propagating scalar degrees of freedom of maximal 
$d=3$ supergravity, are fused into an appropriate generalized vielbein. 
The relevant expressions are found by proceeding from the following 
$\SO$ invariant ansatz for the supersymmetry variations of the 
vector fields in terms of the fermions \Ref{gravitino} and \Ref{chi}
(in a suitable normalization): 
\ba
\d B_\mu{}^m &=& \ft12 e^m_{\,IJ} \, \bar\e^I \psi^J_\mu
        + e^m_{\, A} \, \G^I_{A\dot B} \bar\e^I \gamma_\mu \chi^{\dot B} \\
\d B_{\mu mn} &=& \ft12 e_{mn IJ} \, \bar\e^I \psi^J_\mu +
    e_{mn A} \G^I_{A\dot B} \, \bar\e^I \gamma_\mu  \chi^{\dot B} \la{dBmumn}
\ea

The explicit expressions in a special $SO(16)$ gauge for the new 
bosonic quantities appearing on the r.h.s. of this equation
can be found by comparing the above expressions with the ones obtained
directly from $d=11$ supergravity in the gauge \Ref{vielbein}. It is
already known that \ci{Nico87}
\be
(e^m_{\,IJ},\,e^m_{\,A}):=\left\{
\begin{array}{ll}
\D^{-1}e_a{}^m\G^a_{\a\dot\b}&\text{if }[IJ]\text{ or }A=(\a\dot\b)\\
0                            &\text{otherwise} \la{vielbein1}
\end{array}\right.
\ee
again using the $SO(8)$ decompositions of appendix B. By contrast, 
the objects $\big(e_{mnIJ}\, , \, e_{mnA}\big)$ related to the 
gauge fields $B_{\mu mn}$ -- hence with antisymmetrized {\it lower} 
internal world indices -- have not yet appeared in previous work. 
Matching the r.h.s. of \Ref{dBmumn} with the variations obtained
directly from the $d=11$ supersymmetry variations of \cite{CrJuSc78} 
we find
\be
e_{mn \cA}:= \stackrel{\circ}{e}_{mn\cA} + A_{mnp} e^p_\cA
\ee
where
\be
\stackrel{{ \circ}}{e}_{mn IJ}:=\left\{
               \begin{array}{ll}
\D^{-1}e_m{}^ae_n{}^b (\G_{ab})_{\a\b}&\text{if }[IJ] = [\a\b]\\
- \D^{-1}e_m{}^ae_n{}^b (\G_{ab})_{\dot\a \dot\b}&\text{if }[IJ] 
                         = [{\dot\a \dot \b}] \\
            0   &\text{if } [IJ] = [{\a\dot\b}]  \equiv - [{\dot\b \a}]

\end{array}\right. \la{vielbein2}
\ee
\be
\stackrel{\circ}{e}_{mn A} :=\left\{
              \begin{array}{ll}
4 \D^{-1}e\undersym{{}_{ma} e_{nb}}&\text{if } A=(ab)\\
0                            &\text{if } A=({\a\dot\b})
\end{array}\right. \la{vielbein3}
\ee
Labeling the $\EE$ indices $([IJ],A)$ collectively by
$\cA, \cB, \dots =1,\dots,248$ as in appendix A, the new objects 
$e^m_{\, \cA}$ and $e_{mn \cA}$ together form a rectangular
$(8+28)\times 248$ matrix.

The new vielbeine are manifestly covariant w.r.t. local $\SO$,
thereby enlarging the action of $SO(8)$ of the original theory.
While the supersymmetry variation of $e^m_{\,\cA}$ was already
given in \ci{Nico87}, the variation of $e_{mn\cA}$ has so far not
been determined. In appendix C we show that
\be
\d e_{mnIJ} = -\ft12 \G^{IJ}_{AB} \omega^A e_{mnB} \qquad  
\d e_{mnA} = \ft14 \G^{IJ}_{AB} \omega^B e_{mnIJ}
\ee
with the local $N=16$ supersymmetry parameter
\be
\omega^A := \ft14 \G^I_{A\dot A} \bar\varepsilon^I \chi^{\dot A},
\ee
In deriving this result, a compensating $\SO$ rotation must be 
taken into account to restore the triangular gauge. It is an important 
consistency
check that this compensating rotation comes out to be the 
same for $e^m_{\,\cA}$ and $e_{mn\cA}$, and that the 
resulting transformation is exactly the same as for both fields,
as required by consistency. We can therefore combine the
supersymmetry variations of the generalized vielbeine with
local $\SO$ into the $\EE$ covariant form
\be\label{susy}
\d e^m_{\,\cA} = {f_{\cA\cB}}^\cC \omega^\cB e^m_{\,\cC} \qquad
\d e_{mn\cA} = {f_{\cA\cB}}^\cC \omega^\cB e_{mn \cC}
\ee
We also note that the components $e_{mn\cA}$ were not needed in
\cite{Nico87} because they cannot appear in the supersymmetry
variations of the fermionic fields (the latter depend on the 
3-form potential via the 4-index field strengths only).

All fields transform under general coordinate transformations
in eleven dimensions. Splitting the $d=11$ parameter as 
$\xi^M=(\xi^\mu,\xi^m)$, the transformations generated by $\xi^\mu$
take the standard form. For the remaining ``internal'' coordinate 
transformations with parameter $\xi^m$, we have
\ba\la{gct1}
\d {B_\mu}^m &=& \cD_\mu \xi^m + \xi^n \p_n {B_\mu}^m   \non 
\d B_{\mu mn} &=& 2 \p_{[m} \xi^p B_{\mu n]p} + \xi^p \p_p B_{\mu mn}
\ea
Furthermore,
\ba\la{gct2}
\d e^m_{\, \cA} &=& \xi^p \p_p e^m_{\, \cA} - 
    \p_p \xi^m \,  e^p_{\, \cA} - \p_p \xi^p \, e^m_{\, \cA} \non
\d e_{mn\cA} &=& \xi^p \p_p e_{mn \cA} +
    2\p_{[m} \xi^p \,  e_{n]p \cA} - \p_p \xi^p \, e_{mn \cA}
\ea
Whereas all the $\SO$ fields considered previously were inert
under tensor gauge transformations $\d A_{MNP} =3\p_{[M} \xi_{NP]}$,
the non-invariance of $e_{mn \cA}$ under such transformations, 
due to the appearance of the 3-index field $A_{mnp}$ in its definition, 
is a new feature. Specifically, under tensor gauge transformations 
with parameter $\xi_{mn}$ we have 
\ba\la{tgt1}
\d {B_\mu}^m &=& 0   \non
\d B_{\mu mn}&=& \cD_\mu \xi_{mn} - 2 {B_\mu}^p \p_{[m} \xi_{n]p}
\ea 
and
\ba\la{tgt2}
\d e^m_{\,\cA} &=& 0   \non
\d e_{mn\cA} &=& \p_p \xi_{mn} \, e^p_{\,\cA} 
                   + 2 \p_{[m}\xi_{n]p} \, e^p_{\,\cA} 
\ea
When combined with the previous coordinate transformations, these 
formulas are very suggestive of a unification of the internal
coordinate transformations and tensor gauge transformations,
based on combining the internal coordinate transformation parameters
$\xi^m$ with the residual tensor gauge parameters $\xi_{mn}$
into a single set $(\xi^m,\xi_{np})$ of 36 parameters. In the
remaining sections we will show how these local symmetries are 
related to the the maximal nilpotent commuting subalgebra of $\EE$.

\section{Solution of algebraic constraints on the generalized vielbein}
{}From the expressions \Ref{vielbein1}-\Ref{vielbein3} one can deduce 
a number of algebraic constraints on the generalized vielbein. 
They are \ci{MelNic98} 
\ba
e^m_A e^n_A - \ft12 e^m_{IJ} e^n_{IJ} &=& 0 \la{constr-1}\\[1ex]
\G^{IJ}_{AB}\big(e^m_B e^n_{IJ}-e^n_B e^m_{IJ}\big) &=& 0 \non
\G^{IJ}_{AB} e^m_A e^n_B   +4 e^m_{K[I} e^n_{J]K} &=& 0 \la{constr-2} 
\ea
and
\ba
e^{(m}_{IK} e^{n)}_{JK} -\ft1{16}\d_{IJ}e^m_{KL}e^n_{KL} &=& 0 \non
e^{(m}_{[IJ} e^{n)}_{KL]} +\ft1{24}e^m_A \G^{IJKL}_{AB} e^n_B &=& 0 \non
\G^K_{\dot{A}B}e^{(m}_B e^{n)}_{KL} 
  -\ft1{14}\G^{IKL}_{\dot{A}B} e^{(m}_B e^{n)}_{KL} &=& 0 \la{constr-3}
\ea
For the new components, we find
\ba
e_{mnA} e_{pqA} - \ft12 e_{mnIJ} e_{pqIJ} &=& 0\la{constr-1a} \\[1ex]
\G^{IJ}_{AB}\big(e_{mnB} e_{pqIJ}-e_{pq B} e_{mnIJ}\big) &=& 0 \non
\G^{IJ}_{AB} e_{mnA} e_{pqB}   
   +4 e_{mnK[I} e_{pq J]K} &=& 0 \la{constr-2a} 
\ea
and
\ba
e_{mnA} e^p_{\,A} - \ft12 e_{mnIJ} e^p_{\,IJ} &=& 0 \la{constr-1b}\\[1ex]
\G^{IJ}_{AB}\big(e_{mnB} e^p_{\,IJ}-e^p_{ \,B} e_{mn IJ}\big) &=& 0 \non
\G^{IJ}_{AB} e_{mn A} e^p_{\,B}   
   +4 e_{mn K[I} e^p_{\, J]K} &=& 0 \la{constr-2b} 
\ea
whereas no analog of \Ref{constr-3} exists for the components $e_{mn\cA}$.
All these relations are proved by decomposing the $\SO$ into $SO(8)$
and verifying the vanishing of their components case by case.

The above constraints can be elegantly rewritten in an 
$\EE$ covariant form by means of the projectors onto the 
invariant subspaces of the tensor product of $\EE$ representations 
$\mathbf{248}\otimes\mathbf{248}$ w.r.t.\ the decomposition
\hbox{${\bf 248}\otimes{\bf 248} = {\bf 1}\oplus{\bf 248}\oplus  
 {\bf 3875}\oplus{\bf 27000}\oplus{\bf 30380}$}. The relevant
projectors are explicitly given in terms of $\EE$ structure 
constants by \ci{KoNiSa99}
\ba\la{projind}
(\cP_{1})_{\cA\cB}{}^{\cC\cD}
 &=& \ft{1}{248}\,\eta_{\cA\cB} \eta^{\cC\cD},\\
(\cP_{248})_{\cA\cB}{}^{\cC\cD} 
 &=& -\ft{1}{60}\, f^\cE{}_{\cA\cB} f_\cE{}^{\cC\cD},\non
(\cP_{3875})_{\cA\cB}{}^{\cC\cD} 
 &=& \ft{1}{7}\,  \d_{(\cA}^{\hphantom{(}\cC} \d_{\cB)}^{\cD} 
    -\ft{1}{56}\, \eta_{\cA\cB} \eta^{\cC\cD}
    -\ft{1}{14}\, f^\cE{}_\cA{}^{(\cC} f_{\cE\cB}{}^{\cD)},
\ea
It is straightforward to see that \Ref{constr-1}, \Ref{constr-2}
\Ref{constr-1a}-\Ref{constr-2b} are equivalent to
\ba\la{constr-e8}
(\cP_{j})_{\cA\cB}{}^{\cC\cD} e^m_{\,\cC} e^n_{\,\cD} &=& 0 \non
(\cP_{j})_{\cA\cB}{}^{\cC\cD} e^m_{\, \cC} e_{pq\cD} &=& 0 \non
(\cP_{j})_{\cA\cB}{}^{\cC\cD} e_{mn\cC} e_{pq\cD} &=& 0 
\ea
for $j=1$ and $248$. It takes a little more work to verify that 
\Ref{constr-3} can be expressed in the form\footnote{A further 
relation given in \ci{MelNic98} 
$$
e^{(m}_{\,A} e^{n)}_{\,B} +\ft{1}{32}(\G^{IJ}\G^{KL})_{AB}e^m_{\,IJ}e^n_{\,KL}
-\ft{1}{16}\G^{IJ}_{AC} e^{(m}_{\,C} e^{n)}_{\,D} \G^{KL}_{DB} =0
$$
can be shown to be equivalent to \Ref{constr-3} by means of a Fierz 
transformation and yields no new information. Therefore, there are no 
relations involving the projectors $\cP_{27000}$ and $\cP_{30380}$.}  
\be\la{constr-ee8}
(\cP_{{3875}})_{\cA\cB}{}^{\cC\cD} e^m_{\,\cC} e^n_{\,\cD} = 0
\ee
Observe the invariance of the constraints (\ref{constr-e8})
and (\ref{constr-ee8}) under the combined general coordinate and 
tensor gauge transformations (\ref{gct2}) and (\ref{tgt2}):
the transformed generalized vielbein still obeys all constraints.

We next demonstrate that the $\EE$ invariant algebraic relations 
on the generalized vielbein given above can be solved in terms of 
an $\EE$ matrix $\cV$. For the dimensionally reduced theory this 
is, of course, the expected result \cite{Juli83,MarSch83}), but with
the important difference that the dependence on all eleven 
coordinates is here retained. Thus, $\EE$ is already present in 
eleven dimensions, though not a symmetry of the theory. The 
existence of $\cV$ also clarifies why we end up with the 
right number of physical degrees of freedom, a fact that cannot 
be directly ascertained by counting the constraints: being subject 
to local $\SO$ transformations, the matrix $\cV$ possesses just 
the $128 = 248 -120$ degrees of freedom of the coset $\EE/\SO$. 
Thus, the counting works exactly as for the reduced theory.

To corroborate this claim, consider the $\EE$ Lie algebra 
valued matrices
\ba
\Zt^m &:=&  e^m_{\,\cA} \cX^\cA \equiv\ft12 e^m_{\,IJ} X^{IJ}
           + e^m_{\, A}  Y^A \non
\Zt_{mn} &:=&  e_{mn\,\cA} \cX^\cA \equiv
  \ft12 e_{mn\,IJ} X^{IJ}+e_{mn\,A} Y^A \la{Ztm}
\ea
{}From the relations presented in the foregoing section we infer 
that these matrices  commute (cf. \Ref{constr-2}, \Ref{constr-2a}, 
\Ref{constr-2b})
\be
[ \Zt^m \, , \, \Zt^n ] =
[ \Zt^m \, , \, \Zt_{pq} ] = 
[ \Zt_{mn} \, , \, \Zt_{pq} ] = 0 
\ee 
and are nilpotent (cf. \Ref{constr-1}, \Ref{constr-1a}, \Ref{constr-1b})
\be
{\rm Tr} \Big( \Zt^m \Zt^n \Big) = 
{\rm Tr} \Big( \Zt^m \Zt_{pq} \Big) =
{\rm Tr} \Big( \Zt_{mn} \Zt_{pq} \Big) = 0
\ee
(ensuring that any linear combination of these matrices has norm
zero). Since they are linearly independent (as is most easily 
checked by setting ${e_m}^a = \d_m^a$ in the original definition), 
they form a 36 dimensional abelian nilpotent subalgebra of $\EE$. 
There is only one such algebra, which is unique up to 
conjugation \cite{Malc45,CJLP98}. Consequently, there must
exist an $\EE$ matrix $\cV$ such that
\be
\Zt^m = \cV^{-1} Z^m \cV \qquad \Zt_{mn} = \cV^{-1} Z_{mn} \cV
\ee
where $Z^m$ and $Z_{mn}$ are the 8+28 nilpotent generators of $\EE$
introduced in the appendix. The assignment of the 8+28 vielbein 
components to these generators here is uniquely determined by
$SL(8,\R)$ covariance; its correctness will be confirmed below
when we analyze \Ref{constr-3}. Thus,
\ba\la{eV1}
e^m_{\,\cA} &=& \ft{1}{60} \Tr (Z^m\cV \cX_{\cA} \cV^{-1}) \nn \\
e_{mn\,\cA} &=& \ft{1}{60} \Tr (Z_{mn}\cV \cX_{\cA} \cV^{-1}) 
\ea
By use of the relation
\be
\ft1{60} \Tr \Big(\cX^\cM \cV \cX_\cA \cV^{-1} \Big) = {\cV^\cM}_\cA
\ee
for the adjoint representation we can write 
\be\la{VMA}
e^m_{\,\cA} = {\cV^m}_\cA \qquad  e_{mn\,\cA} = \cV_{mn\,\cA}
\ee
whence the generalized vielbein $(e^m_{\,\cA} , e_{mn\cA})$ is actually 
a rectangular submatrix of $\cV$. Let us emphasize once more that 
$\cV$ still depends on eleven coordinates. 

At this point, a remark concerning our use of indices is in order. 
In the appendix, we use ``flat'' indices $a,b,c =1,\dots ,8$ to 
label the $\EE$ generators in the $SL(8,\R)$ decomposition.
On the other hand, the index $m$ appearing on the l.h.s. of 
\Ref{vielbein1} should be viewed as ``curved'' in the sense that 
it is acted on by internal coordinate transformations. 
Of course, there is no need for such a distinction in a flat
background characterized by $e_m{}^a=\d_m^{\,a}$ and $\cV = {\bf 1}$, 
whereas the two kinds of indices no longer coincide for curved 
backgrounds characterized by non-trivial $e_m{}^a$ and $A_{mnp}$.
The nilpotent generators in \Ref{Ztm} thus represent ``curved'' 
analogs of the ``flat'' generators $Z^a$ and $Z_{ab}$, and the
above relations tell us is that the transition from flat to curved 
configurations is entirely accounted for by the $\EE$ matrix $\cV$. 
This illustrates the enlargement of the geometry in comparison 
with the conventional description $d=11$ supergravity, where the 
achtbein can only be deformed with a $GL(8,\R)$ matrix.

To confirm the consistency of the above solution let us analyze the 
third set of constraints \Ref{constr-3}, which we have not yet discussed. 
Inspection reveals that the desired relation is equivalent to 
\be
\cP_{3875} \big( \Zt^m \otimes \Zt^n \big) = 0
\ee
Making use of the $\EE$ invariance of $\cP_{3875}$, we can replace
curved by flat indices in this relation.
This yields
\be\la{3875}
{(\cP_{3875})_{\cA\cB}}^{cd} \equiv 
 \ft{1}{7}\,  \d_{\cA}^{\hphantom{(}(c} \d_{\cB}^{\, d)} 
 -\ft{1}{56}\, \eta_{\cA\cB} \eta^{cd}
 -\ft{1}{14}\, f^\cE{}_\cA{}^{(c} f_{\cE\cB}{}^{d)} = 0
\ee
By nilpotency, we have $\eta^{cd} =0$, and contracting the
remaining terms with $\cX^\cB$ we see that \Ref{vielbein1} does 
satisfy \Ref{constr-3}, provided the following relation holds for 
all $\cA$
\be
\left[ \left[\cX_\cA, Z^c\right],Z^d\right] 
     = -2 \d_\cA^{(c} Z^{d)}      \la{ZZZ}
\ee
Since the algebra preserves the grading, the relation is trivially
satisfied for all generators except $\cX_\cA = Z_a$, which must be
checked separately. A quick calculation, using the commutation 
relations listed in appendix B, shows that the required relation 
is indeed satisfied. Let us emphasize once more that there is no 
analog of \Ref{3875} for nilpotent elements conjugate to the $Z_{mn}$.

Having established the consistency of \Ref{eV1} it remains to
investigate its uniqueness. It is easy to see that $\cV$ is, in
fact, {\it not} unique because \Ref{eV1} remains unchanged under
the transformation
\be
\cV \longrightarrow n \cdot \cV \qquad {\rm for} \qquad n\in\cN
\ee
where $\cN$ is the Borel subgroup of $\EE$ generated by $Z^m$ and
$Z_{mn}$. This remaining non-uniqueness can be fixed by invoking 
the differential relations
\ba\la{diffrel}
e^m_{\,A} P_\mu^{\, A} &=& {\varepsilon_\mu}^{\nu\rho} {{\cB}_{\nu\rho}}^m \\
e_{mn A} P_\mu^{\, A} &=& {\varepsilon_\mu}^{\nu\rho} {\cB}_{\nu\rho mn}
\ea
where ${{\cB}_{\mu\nu}}^m$ was already defined in \Ref{Bmunu}, and
\be
\cB_{\mu\nu mn} := \cD_\mu B_{\nu mn} -\cD_\nu B_{\mu mn}
      + 4 \p_{[m} {B_{[\mu}}^p B_{\nu]n]p} 
      + 2 \p_{[m} B_{n]\mu\nu}       \la{Bmumn}
\ee
\Ref{diffrel} thus relates the field strengths to (part of) 
the $\EE$ connection $P_\mu^{\, A}$, whose explicit expression 
in terms of the spin connection and the 4-index field strength 
is given in formula (22) of \cite{Nico87}. In the reduction 
to three dimensions these differential constraints 
become the linear duality relations that allow us to trade the 
36 vector fields for their dual scalars. 

\mathversion{bold}
\section{Outlook}
\mathversion{normal}
The matrix $\cV$ plays a role similar to the achtbein ${e_m}^a$,
but also incorporates the tensor degrees of freedom from $A_{mnp}$,
as well as the vector fields ${B_\mu}^m$ and $B_{\mu mn}$. We would now 
like to argue that $\cV \in \EE/\SO$ really {\it is} the appropriate 
vielbein encompassing the propagating degrees of freedom of $d=11$ 
supergravity, in the same way as the ordinary vielbein, viewed as an 
element of $GL(d,\R)/SO(d)$, describes the graviton degrees of freedom
(with a Euclidean signature). To this aim, let us first recall that 
the internal part of the (inverse densitized) metric is recovered 
from the generalized vielbein via the $\SO$ invariant formula 
\be\la{metric1}
\D^{-2} g^{mn} = \ft1{60}e^m_{\,A} e^n_{\,A} 
  = \ft1{120}\left( e^m_{\,A} e^n_{\, A} + 
    \ft12 e^m_{\,IJ} e^n_{\, IJ} \right)
\ee
where the constraint \Ref{constr-1} was used. The summation on the 
r.h.s. breaks $\EE$ to $\SO$ because of the plus sign in front of the 
second term; with a minus sign, the expression would vanish by
the constraints.

Just as for the standard vielbein, and having already introduced
this terminology in the foregoing section, we now interpret the 
indices $\cM,\cN,\dots$ and $\cA,\cB,\dots$ appearing in \Ref{VMA} as 
``curved'' and ``flat'', respectively. This then immediately suggest 
the the following generalization of \Ref{metric1}
\be\la{metric2}
\cG^{\cM\cN} := \ft1{120} {\cV^\cM}_\cA {\cV^\cN}_\cA
     \equiv \ft1{120} \left( {\cV^\cM}_A {\cV^\cN}_A +
     \ft12 {\cV^\cM}_{IJ} {\cV^\cN}_{IJ} \right)
\ee
By construction, this metric is invariant under local $\SO$, 
which acts as
\be
{\cV^\cM}_\cA \longrightarrow {\cV^\cM}_\cB {\Sigma^\cB}_\cA 
\ee
on the 248-bein with $\Sigma$ in the ${\bf 120} \oplus {\bf 128}$ 
representation of $\SO$ (and depending on all eleven coordinates).
As we showed before, a ``bosonized'' version of local supersymmetry
formally extends this to an action of a full local $\EE$ acting
from the right on ${\cV^\cM}_\cA$ (which, however, does not leave
$\cG^{\cM\cN}$ inert any more).

Similarly, we now have a combined action of the internal coordinate
and tensor gauge transformations on the 248-bein ${\cV^\cM}_\cA$,
which is analogous to the action of general coordinate
transformations on the standard vielbein. Namely, from the previous 
formulas \Ref{gct2} and \Ref{tgt2} we can directly read off their 
action on the $36\times 248$ submatrix of ${\cV^\cM}_{\cA}$:
\ba
\d \cV^m_{\, \cA} &=& \xi^p \p_p \cV^m_{\, \cA} - 
    \p_p \xi^m \,  \cV^p_{\, \cA} - \p_p \xi^p \, \cV^m_{\, \cA} \non
\d \cV_{mn\,\cA} &=& \xi^p \p_p \cV_{mn\, \cA} - 
    2\p_{[m} \xi^p \,  \cV_{n]p\, \cA} - \p_p \xi^p \, \cV_{mn\, \cA}
\ea
and
\ba
\d \cV^m_{\,\cA} &=& 0   \non
\d \cV_{mn\,\cA} &=& \p_p \xi_{mn} \, \cV^p_{\,\cA} 
                   + 2 \p_{[m}\xi_{n]p} \, \cV^p_{\,\cA} 
\ea
This action can be extended to the full 248-bein via
\ba
\d \cV &=& \xi^p \p_p  \cV +
    \p_p \xi^q \left( {E_q}^p - \ft38 \d^p_q N \right) \cV  \non
\d \cV &=& \p_p \xi_{qr} E^{pqr} \, \cV
\ea
by means of the $\EE$ Lie algebra matrices given in appendix A
(although this is usually not done, the general coordinate 
transformations on the standard vielbein can be cast into an 
analogous form by use of $GL(d,\R)$ matrices). It is important 
here that this action manifestly preserves $\EE$: the transformed
vielbein is still an element of $\EE$. It is also straightforward
to exponentiate the infinitesimal action to a full ``diffeomorphism''
generated by the pair $(\xi^m, \xi_{mn})$. Intriguingly, the 
missing ``transport term'' with $\xi_{mn}$ suggests a hidden 
dependence on 28 further coordinates $x_{mn}$, but it remains
to be seen whether such an extension exists. Finally, it is clear
that the rigid $\EE$ invariance of the dimensionally reduced theory 
emerges from the above local symmetries in the same way as rigid 
$GL(d,\R)$ symmetry emerges in the torus reduction as a remnant 
of general coordinate invariance.

\newpage
\begin{appendix}
\renewcommand{\theequation}{\Alph{section}.\arabic{equation}}
\renewcommand{\thesection}{Appendix \Alph{section}:}
\mathversion{bold}
\section{The $\SO$ decomposition of $\EE$.}
\mathversion{normal}

In the standard $SO(16)$ decomposition
${\bf 248}\ra {\bf120}\oplus {\bf 128}$, the $\EE$ Lie algebra 
generators are $\cX^\cA = (X^{IJ},Y^A)$, 
with $SO(16)$ vector and spinor indices $I,J =1,\ldots,16$ and 
$A=1,\ldots,128$, respectively. They obey
\ba
\big[X^{IJ},X^{KL}\big] &=& 4\, \d\oversym{{}^{I[K}X^{L]J}}\non
\big[X^{IJ},Y^{A} \big] &=&  \ft14 \G^{IJ}_{AB}X^{IJ}\non
\big[Y^{A} ,Y^{B} \big] &=& -\ft12 \G^{IJ}_{AB}Y^{B}\la{E8}
\ea
The totally antisymmetric $\E$ structure constants $f^{\cA\cB\cC}$
therefore possess the non-vanishing components
\be
f^{I\!J,\,K\!L,\,M\!N} = 
-8\, \d\!\oversym{^{I[K}_{\vphantom{M}}\,\d_{MN}^{L]J}},\qquad
f^{I\!J,\,A,\,B}   = -\ft12 \G^{IJ}_{AB} \la{fABC}
\ee
The Cartan-Killing form is given by
\be
\eta^{\cA\cB}= \frac1{60} \tr \cX^\cA \cX^\cB 
             = -\frac1{60} f^\cA{}_{\cC\cD}f^{\cB\cC\cD}
\ee
with components $\eta^{AB}=\d^{AB}$ and $\eta^{I\!J\,K\!L}=-2\d^{IJ}_{KL}$.
When summing over antisymmetrized index pairs $[IJ]$, an extra factor 
of $\frac12$ is always understood. 

\mathversion{bold}
\section{The $SL(8,\R)$ decomposition of $\EE$.}
\mathversion{normal}
To recover the $SL(8,\R)$ basis of \ci{CJLP98}, we will further decompose 
the above representations into representations of the subgroup
$SO(8)\equiv \big( SO(8)\times SO(8)\big)_{\rm diag}\subset SO(16)$.
The indices corresponding to the ${\bf 8}_v , {\bf 8}_s$ and
${\bf 8}_c$ representations of $SO(8)$, respectively, will be denoted by
$a$, $\a$ and $\dot\a$. After a triality rotation the $\SO$
vector and spinor representations decompose as 
\ba\la{SO8dec}
\bf{16}  &\lra&   \bf{8_s}\oplus\bf{8_c}\non 
\bf{128}_s  &\lra&  (\bf{8_s}\!\otimes\!\bf{8_c}) \oplus
(\bf{8_v}\!\otimes\!\bf{8_v}) = {\bf 8}_v \oplus {\bf 56}_v \oplus
{\bf 1} \oplus {\bf 28} \oplus {\bf 35}_v \non
\bf{128}_c &\lra & (\bf{8_v}\!\otimes\!\bf{8_s}) \oplus
(\bf{8_c}\!\otimes\!\bf{8_v}) = {\bf 8}_s \oplus {\bf 56}_s \oplus
                                {\bf 8}_c \oplus {\bf 56}_c  ,
\ea
respectively. We thus have $I=(\a,\dot\a)$ and $A=(\a\dot\b,ab)$, 
and the $\E$ generators decompose as 
\be
X^{[IJ]}\ra (X^{[\a\b]},X^{[\dot\a\dot\b]},X^{\a\dot\b})\;,
 \quad Y^A \ra (Y^{\a\dot\a}, Y^{ab}) \la{SO(8)} \;. 
\ee
Next we regroup these generators as follows. The 63 generators 
\ben
E_a{}^b :=  \ft18 (\G^{ab}_{\a\b}X^{[\a\b]} 
                    +\G^{ab}_{\dot\a\dot\b}X^{[\dot\a\dot\b]})
             +Y^{(ab)} -\ft18 \d^{ab} Y^{cc} \,,\nn 
\een
for $1\le a,b\le 8$ span an $SL(8,\R)$ subalgebra of $\EE$.
The generator 
\ben
N := Y^{cc}
\een
extends this subalgebra to $GL(8,\R)$. The remainder of the $\EE$ Lie 
algebra then decomposes into the following representations of $SL(8,\R)$:
\ba
Z^a     &:=& \ft14 \G^a_{\a\dot\a}(X^{\a\dot\a}+Y^{\a\dot\a}) \,,\non
Z_{ab}  &:=& \ft18 \left(\G^{ab}_{\a\b}X^{[\a\b]} 
                   -\G^{ab}_{\dot\a\dot\b}X^{[\dot\a\dot\b]}\right)
             +Y^{[ab]} \,,\non 
E^{abc} &:=& -\ft14 \G^{abc}_{\a\dot\a} (X^{\a\dot\a}-Y^{\a\dot\a}) \la{+}\,
\ea
and
\ba
Z_a     &:=& -\ft14 \G^a_{\a\dot\a}(X^{\a\dot\a}-Y^{\a\dot\a}) \,, \non
Z^{ab}  &:=&  -\ft18 (\G^{ab}_{\a\b}X^{[\a\b]} 
                    -\G^{ab}_{\dot\a\dot\b}X^{[\dot\a\dot\b]})
              +Y^{[ab]} \,, \non
E_{abc} &:=&  \ft14 \G^{abc}_{\a\dot\a} (X^{\a\dot\a}+Y^{\a\dot\a}) \la{-}\,, 
\ea
The Cartan subalgebra is spanned by the diagonal elements 
$E_1{}^1,\dots, E_7{}^7$ and $N$, or, equivalently, 
by $Y^{11}, \dots , Y^{88}$. Obviously, the elements $E_a{}^b$ 
for $a<b$ (or $a>b$) together with the elements \Ref{+} (or \Ref{-}) 
for $a<b<c$ generate the Borel subalgebra of $\EE$ associated 
with the positive (negative) roots of $\EE$. Furthermore, these 
generators are graded w.r.t. the number of times the root 
$\a_8$ (corresponding to the element $N$ in the Cartan subalgebra)
appears, such that for any basis generator $X$ we have
$[N, X] = {\rm deg}(X)\cdot X$. The degree can be read off from 
\ben\begin{array}{rclrcl}
{}[N,Z^a]  &\!\!\!\!=\!\!\!\!& 3\, Z^a &
{}[N,Z_a]  &\!\!\!\!=\!\!\!\!& -3\, Z_a \\[1ex]
{}[N,Z_{ab}]  &\!\!\!\!=\!\!\!\!& 2\, Z_{ab} &
{}[N,Z^{ab}]  &\!\!\!\!=\!\!\!\!& -2\, Z^{ab} \\[1ex]
{}[N,E^{abc}]  &\!\!\!\!=\!\!\!\!&  E^{abc} &
{}[N,E_{abc}]  &\!\!\!\!=\!\!\!\!& - E_{abc} \\[1ex]
{}[N,E_a{}^b]  &\!\!\!\!=\!\!\!\!& 0 &&&
\end{array}\een
The remaining commutation relations are given by
\ben\begin{array}{rclrcl}
{}[Z^a,Z^b] &\!\!\!\!=\!\!\!\!& 0 &
{}[Z_a,Z_b] &\!\!\!\!=\!\!\!\!& 0           \\[1ex]
{}[Z_a,Z^b] &\!\!\!\!=\!\!\!\!& E_a{}^b - \ft38 \d_a^b N &
\end{array}\een

\ben\begin{array}{rclrcl}
{}[Z_{ab},Z^c] &\!\!\!\!=\!\!\!\!& 0&
{}[Z_{ab},Z_c] &\!\!\!\!=\!\!\!\!& -E_{abc}\\[1ex]
{}[Z_{ab},Z_{cd}] &\!\!\!\!=\!\!\!\!& 0&
{}[Z_{ab},Z^{cd}] &\!\!\!\!=\!\!\!\!& 
  4\d_{[a}^{[c} E_{b]}^{\vphantom{[}}{}^{d]}_{\vphantom{]}} 
  +\ft12 \d_{ab}^{cd} N\\[1ex]
{}[Z^{ab},Z^c] &\!\!\!\!=\!\!\!\!& -E^{abc}&
{}[Z^{ab},Z_c] &\!\!\!\!=\!\!\!\!& 0
\end{array}\een

\ben\begin{array}{rclrcl}
{}[E^{abc},Z^d] &\!\!\!\!=\!\!\!\!& 0&
{}[E_{abc},Z^d] &\!\!\!\!=\!\!\!\!& 3 \d^d_{[a} Z_{bc]}^{\vphantom{a}}\\[1ex]
{}[E^{abc},Z_{de}] &\!\!\!\!=\!\!\!\!& -6\d^{[ab}_{de}Z^{c]}_{\vphantom{d}}&
{}[E_{abc},Z_{de}] &\!\!\!\!=\!\!\!\!& 0\\[1ex]
{}[E^{abc},E^{def}] &\!\!\!\!=\!\!\!\!& -\ft{1}{32}\e^{abcdefgh}Z_{gh}&
{}[E_{abc},E_{def}] &\!\!\!\!=\!\!\!\!& \ft{1}{32}\e_{abcdefgh}Z^{gh}\\[1ex]
{}[E^{abc},Z_d] &\!\!\!\!=\!\!\!\!& 3 \d_d^{[a} Z^{bc]}_{\vphantom{d}}&
{}[E_{abc},Z_d] &\!\!\!\!=\!\!\!\!& 0\\[1ex]
{}[E^{abc},Z^{de}] &\!\!\!\!=\!\!\!\!& 0&
{}[E_{abc},Z^{de}] &\!\!\!\!=\!\!\!\!& 
  6\d^{de}_{[ab}Z_{c]}^{\vphantom{d}}\\[1ex]
{}[E^{abc},E_{def}] &\!\!\!\!=\!\!\!\!& -18 \d^{[ab}_{[de} {E_{f]}}^{c]} 
  -\ft34 \d^{abc}_{def} N
&&&
\end{array}\een

\ben\begin{array}{rclrcl}
{}[E_a{}^b,Z^c] &\!\!\!\!=\!\!\!\!& -\d_a^c Z^b +\ft18 \d_a^b Z^c&
{}[E_a{}^b,Z_c] &\!\!\!\!=\!\!\!\!&  \d_c^b Z_a -\ft18 \d_a^b Z_c\\[1ex]
{}[E_a{}^b,Z_{cd}]&\!\!\!\!=\!\!\!\!&
  -2\d^b_{[c}Z_{d]a}^{\vphantom{a}}-\ft14 \d_a^b Z_{cd}&
{}[E_a{}^b,Z^{cd}]&\!\!\!\!=\!\!\!\!&
  2 \d_a^{[c}Z^{d]b}_{\vphantom{a}} +\ft14 \d_a^b Z^{cd}\\[1ex]
{}[E_a{}^b,E^{cde}]&\!\!\!\!=\!\!\!\!&
  -3 \d_a^{[c} E^{de]b}_{\vphantom{a}} +\ft38 \d_a^b E^{cde}&
{}[E_a{}^b,E_{cde}]&\!\!\!\!=\!\!\!\!&
  3\d^b_{[c} E_{de]a}^{\vphantom{a}} -\ft38 \d_a^b E_{cde}\\[1ex]
{}[E_a{}^b,E_c{}^d] &\!\!\!\!=\!\!\!\!& \d^b_c E_a{}^d -\d_a^d E_c{}^b&
{}&&\\
\end{array}\een
The elements $\{ Z^a, Z_{ab} \}$ (or equivalently 
$\{ Z_a, Z^{ab} \}$) span the maximal 36-dimensional abelian
nilpotent subalgebra of $\EE$ \ci{Malc45,CJLP98}. 

Finally, the generators are normalized according to
\ba
\tr (N N)             &=& 60 \!\cdot\!8\,, \non
\tr (Z^a Z_b)         &=& 60 \, \d^a_b \,, \non
\tr (Z^{ab} Z_{cd})   &=& 60 \!\cdot\! 2!\, \d^{ab}_{cd} \,, \non
\tr (E_{abc} E^{def}) &=& 60 \!\cdot\! 3!\, \d_{abc}^{def} \,,\non
\tr (E_a{}^b E_c{}^d) &=& 60 \, \d_a^d \d^b_c
                             - \ft{15}2 \d^b_a \d_c^d
 \,,\nn
\ea
with all other traces vanishing.

\section{Supersymmetry variations of the generalized vielbein}

The supersymmetry variations of the achtbein and the relevant
components of the 3-form potential with our normalization 
read as follows in the $SO(8)$ basis:
\ba
\d e_m{}^a &=& \ft12\,\left(\bar\varepsilon_{\alpha}\,
\G^a_{\a\dot\b}\Psi_{m\dot\b}-\bar\varepsilon_{\dot\alpha}\,
\bar{\G}^a_{\dot\a\b}\Psi_{m\b}\right)\;, \label{susyE}\\
\d A_{mnp} &=& -\ft32\,\left(
\bar\varepsilon_{\alpha}\,(\G_{[mn})_{\a\b}\Psi_{p]}{}_{\b}-
\bar\varepsilon_{\dot\alpha}\,
(\bar{\G}_{[mn})_{\dot\a\dot\b}\Psi_{p]}{}_{\dot\b}\right) \;.
\nn 
\ea
(The relative minus signs in the second terms on the r.h.s. are due
the fact that the Dirac conjugate spinors are appropriate to $d=3$ 
and differ from the ones in $d=11$ by an extra factor $\G^9$). 
These formulas must now be compared with the $\SO$ covariant
ones in terms of the generalized vielbein. The latter are most
conveniently computed in terms of the matrices $\Zt^m$, $\Zt_{mn}$
from \Ref{Ztm}, making use of the $SL(8,\R)$ decomposition of $\EE$ 
described in the appendix B. In the special $SO(16)$ gauge \Ref{vielbein1},
\Ref{vielbein2}, \Ref{vielbein3} these matrices take the form
\ba
\Zt^m &=&  e^m{}_{\!\cA} \,\cX^\cA ~=~ 4\,\D^{-1} e_a{}^m\,Z^a\;,
\label{gv}\\[1ex]
\Zt_{mn} &=&  e_{mn\,\cA} \,\cX^\cA ~=~ 
4\,\D^{-1} e_m{}^a\,e_n{}^b\,Z_{ab} + A_{mnp}\,\Zt^p \;,\nn
\ea
in the upper Borel subalgebra. The $\SO$ covariant supersymmetry
variations have been presented in \Ref{susy} and can be written as 
\be\label{susy2}
\d\,\Zt^m =  \left[\Zt^m\,, \,\omega\right] \;,  \qquad
\d\,\Zt_{mn} = \left[\Zt_{mn}\,, \,\omega\right]  \;,
\ee
with $\omega$ given by
\be
\omega~:=~ \ft14 \left(\G^I_{A\dot A} \bar\varepsilon^I \chi^{\dot A}\,Y^A
+\ft12\, \omega^{IJ}_{\rm comp}\,X^{IJ} \right)  \;,
\ee
where $\omega^{IJ}_{\rm comp}\,X^{IJ}$ is the compensating $SO(16)$
rotation to restore the triangular gauge of \Ref{gv}. Upon
decomposition into the $SO(8)$ fields, the first term yields
\ba
\G^I_{A\dot A} \bar\varepsilon^I \chi^{\dot A}\,Y^A
&=& 2\left(\bar\varepsilon_{\a}\Psi_{a\a}+
\bar\varepsilon_{\dot\a}\Psi_{a\dot\a}\right)\,(Z^a\!+\!Z_a) 
\label{omega11}\\
&&{}
+\left(\bar\varepsilon_{\a}\G^{ab}_{\a\b}\Psi_{b\b}+
\bar\varepsilon_{\dot\a}\bar{\G}^{ab}_{\dot\a\dot\b}\Psi_{b\dot\b}
\right)\,(Z^a\!+\!Z_a)\non
&&{}-\ft12\,\left(\bar\varepsilon_{\a} \G^{abc}_{\a\dot\b}
\,\Psi_{c\dot\b}+\bar\varepsilon_{\dot\a} \bar{\G}^{abc}_{\dot\a\b}
\,\Psi_{c\b}\right) \,(Z_{ab}\!+\!Z^{ab}) \non
&&{}-\ft12\,\left(\bar\varepsilon_{\a} \G^{ab}_{\a\b} 
\,\Psi_{c\b} -
\bar\varepsilon_{\dot\a} \bar{\G}^{ab}_{\dot\a\dot\b} 
\,\Psi_{c\dot\b} \right)\,(E^{abc}\!+\!E_{abc}) \non
&&{}-\left(\bar\varepsilon_{\alpha}\G^a_{\a\dot\b}\,
\Psi_{b\dot\b} -
\bar\varepsilon_{\dot\a}\bar{\G}^a_{\dot\a\b}\,
\Psi_{b\b}\right)
\,\left(E_a{}^b\!+\!E_b{}^a\!-\!\ft34\d^{ab}\,N\right)
\;,
\nn
\ea
whereas $\omega^{IJ}_{\rm comp}\,X^{IJ}$ is determined in such a way
as to rotate $\omega$ back into the upper Borel subalgebra. The
resulting $\omega$ will then preserve the special gauge choice
\Ref{gv}. Explicitly, it takes the form
\ba
\omega&\equiv&
\G^I_{A\dot A} \bar\varepsilon^I \chi^{\dot A}\,Y^A
+\ft12 \omega^{IJ}_{\rm comp}\,X^{IJ}  \label{omega1}\\[1ex] 
&=&
4\left(\bar\varepsilon_{\a}\Psi_{a\a}+
\bar\varepsilon_{\dot\a}\Psi_{a\dot\a}\right)\,Z^a
+2\left(\bar\varepsilon_{\a}\G^{ab}_{\a\b}\Psi_{b\b}+
\bar\varepsilon_{\dot\a}\bar{\G}^{ab}_{\dot\a\dot\b}\Psi_{b\dot\b}
\right)\,Z^a\non
&&{}-\left(\bar\varepsilon_{\a} \G^{abc}_{\a\dot\b}
\,\Psi_{c\dot\b}+\bar\varepsilon_{\dot\a} \bar{\G}^{abc}_{\dot\a\b}
\,\Psi_{c\b}\right) \,Z_{ab} \non
&&{}-\left(\bar\varepsilon_{\a} \G^{ab}_{\a\b} 
\,\Psi_{c\b} -
\bar\varepsilon_{\dot\a} \bar{\G}^{ab}_{\dot\a\dot\b} 
\,\Psi_{c\dot\b} \right)\,E^{abc} \non
&&{}-\left(\bar\varepsilon_{\alpha}\G^a_{\a\dot\b}\,
\Psi_{b\dot\b} -
\bar\varepsilon_{\dot\a}\bar{\G}^a_{\dot\a\b}\,
\Psi_{b\b}\right) \,\left(2E_a{}^b\!-\!\ft34\d^{ab}\,N\right)
\nn
\ea
{}From \Ref{susy2} we then find the supersymmetry variation:
\ba
\d\,\Zt^m &=& -2\D^{-1}\,Z^a\,
\left(\bar\varepsilon_{\alpha}\G^m_{\a\dot\b}\,
\Psi_{a\dot\b}-\bar\varepsilon_{\dot\a}\bar\G^m_{\dot\a\b}\,
\Psi_{a\b}\right)\\
&&{}-2\D^{-1}Z^a\,e_a{}^m
\left(\bar\varepsilon_{\alpha}\G^b_{\a\dot\b}\,
\Psi_{b\dot\b}-\bar\varepsilon_{\dot\a}\bar\G^b_{\dot\a\b}\,
\Psi_{b\b}\right)
\non[2ex]
\d\,\Zt_{mn} &=& 4\D^{-1}\,Z_{ab}\,
e_m{}^a\,\left(\bar\varepsilon_{\alpha}\,\G^b_
{\a\dot\b}\Psi_{n\dot\b}-
\bar\varepsilon_{\dot\alpha}\,\bar{\G}^b_{\dot\a\b}\Psi_{n\b}\right)\\
&&{}-2\D^{-1}\,Z_{ab}\,e_m{}^ae_n{}^b\,
\left(\bar\varepsilon_{\alpha}\,
\G^c_{\a\dot\b} \Psi_{c\dot\b}-\bar\varepsilon_{\dot\alpha}\,
\bar{\G}^c_{\dot\a\b} \Psi_{c\b}\right) \non
&&{}
-\ft32\,\Zt^p\,\left(
\bar\varepsilon_{\alpha}\,(\G_{[mn})_{\a\b}\Psi_{p]}{}_{\b}-
\bar\varepsilon_{\dot\alpha}\,
(\bar{\G}_{[mn})_{\dot\a\dot\b}\Psi_{p]}{}_{\dot\b}\right) \non
&&{}+A_{mnp}\,\d\Zt^p 
\nn
\ea
and thus agreement with \Ref{susyE}, since the one-but-last term 
is just $\d A_{mnp} \Zt^p$.

\end{appendix}


\begin{thebibliography}{10}

\bibitem{CrJuSc78}
E.~Cremmer, B.~Julia, and J.~Scherk.
\newblock {\em Phys. Lett.\/} {\bf 76B} (1978) 409.

\bibitem{CreJul79}
E.~Cremmer and B.~Julia.
\newblock {\em Nucl. Phys.\/} {\bf B159} (1979) 141.

\bibitem{WitNic86}
B.~de~Wit and H.~Nicolai.
\newblock {\em Nucl. Phys.\/} {\bf B274} (1986) 363.

\bibitem{Nico87}
H.~Nicolai.
\newblock {\em Phys. Lett.\/} {\bf B187} (1987) 316.

\bibitem{WitNic84}
B.~de Wit and H.~Nicolai.
\newblock {\em Nucl. Phys.\/} {\bf B243} (1984) 91

\bibitem{WitNic87}
B.~de~Wit and H.~Nicolai.
\newblock {\em Nucl. Phys.\/} {\bf B281} (1987) 211.

\bibitem{Juli83}
B.~Julia.
\newblock In V.~D. Sabbata and E.~Schmutzer (eds.), {\em Unified field theories
  in more than 4 dimensions\/}, 215--236 (World Scientific, Singapore, 1983).

\bibitem{MarSch83}
N.~Marcus and J.~H. Schwarz.
\newblock {\em Nucl. Phys.\/} {\bf B228} (1983) 145.

\bibitem{Mizo97}
S.~Mizoguchi.
\newblock {\em Nucl. Phys.\/} {\bf B528} (1998) 238.

\bibitem{CJLP98}
E.~Cremmer, B.~Julia, H.~Lu, and C.~N. Pope.
\newblock {\em Nucl. Phys.\/} {\bf B523} (1998) 73.

\bibitem{Malc45}
A.~Malcev.
\newblock {\em Izv. Akad. Nauk SSR, Ser. Mat.\/} {\bf 9} (1945) 291.

\bibitem{MelNic98}
S.~Melosch and H.~Nicolai.
\newblock {\em Phys. Lett.\/} {\bf B416} (1998) 91.

\bibitem{KoNiSa99}
K.~Koepsell, H.~Nicolai, and H.~Samtleben.
\newblock {\em JHEP\/} {\bf 04} (1999) 023.


\end{thebibliography}
\end{document}